\documentclass[aps,pre,twocolumn,showpacs,groupedaddress,amssymb]{revtex4}

\usepackage{graphicx} 

\begin{document}
\title{Anisotropic shear melting and recrystallization \\ of a two-dimensional complex (dusty) plasma}

\author{V. Nosenko}
\email{nosenko@mpe.mpg.de}
\author{A. V. Ivlev}
\author{G. E. Morfill}
\affiliation{Max-Planck-Institut f\"{u}r extraterrestrische Physik, D-85741 Garching, Germany}

\date{\today}
\begin{abstract}
A two-dimensional plasma crystal was melted by suddenly applying localized shear stress. A stripe of particles in the crystal was pushed by the radiation pressure force of a laser beam. We found that the response of the plasma crystal to stress and the eventual shear melting depended strongly on the crystal's angular orientation relative to the laser beam. Shear stress and strain rate were measured, from which the spatially resolved shear viscosity was calculated. The latter was shown to have minima in the regions with high velocity shear, thus demonstrating shear thinning. Shear-induced reordering was observed in the steady-state flow, where particles formed strings aligned in the flow direction.
\end{abstract}
 \maketitle

\section {Introduction}

Shear melting is a non-equilibrium process where a crystalline lattice is ruptured and eventually melted by an external shearing force. Shear melting and related phenomena were experimentally studied in soft condensed matter systems such as colloids \cite{Ackerson:81,Stevens:91,New_book} and complex (dusty) plasmas \cite{Nosenko:04PRL_visc,Feng:10,Nosenko:11PRL,New_book}.

A complex, or dusty plasma is a suspension of small solid particles in an ionized gas \cite{New_book,ShuklaBook}. Particles get charged by collecting electrons and ions from the ambient plasma and interact with each other via a screened Coulomb pair potential. When these interparticle forces are balanced by other forces (gravity, neutral and ion drag, ambipolar electric fields), the particles can form regular structures called plasma crystals. In the presence of gravity, a single-layer (two-dimensional, 2D) plasma crystal can form \cite{Nosenko:04PRL_visc,Feng:10,Nosenko:11PRL}. In a 2D plasma crystal, the in-plane interparticle pair potential can be well approximated by the Yukawa potential. The shear modulus of such crystals is on the scale of $10^{-13}$~N/mm \cite{Nosenko:11PRL}, and therefore they can be easily manipulated, e.g. by the radiation pressure force of a focused laser beam \cite{Chan:04,Nosenko:09PRL}.

Using plasma crystals as model systems, shear melting was studied experimentally in Refs.~\cite{Nosenko:04PRL_visc,Feng:10,Nosenko:11PRL}. Proliferation of supersonic (for shear sound) dislocations was suggested as a possible mechanism of shear melting \cite{Nosenko:11PRL}. Shear flows in melted complex plasmas and their transport properties, e.g shear viscosity were studied in Refs.~\cite{Nosenko:04PRL_visc,Gavrikov:05,Ivlev:07}. One important aspect of shear melting that has not been studied so far is the role of relative angular orientation of the crystal and applied shearing force.

In this work, we studied the response of a 2D complex plasma crystal to a suddenly applied shear stress. The crystal was pre-oriented to align the closely packed rows of particles at $0^{\circ}$ or $30^{\circ}$ (two principal orientations) relative to the shearing force direction. The force was applied by the radiation pressure of a focused laser beam.

\section {Experimental method}

\begin{figure}
\centering
\includegraphics[width=0.85\linewidth]{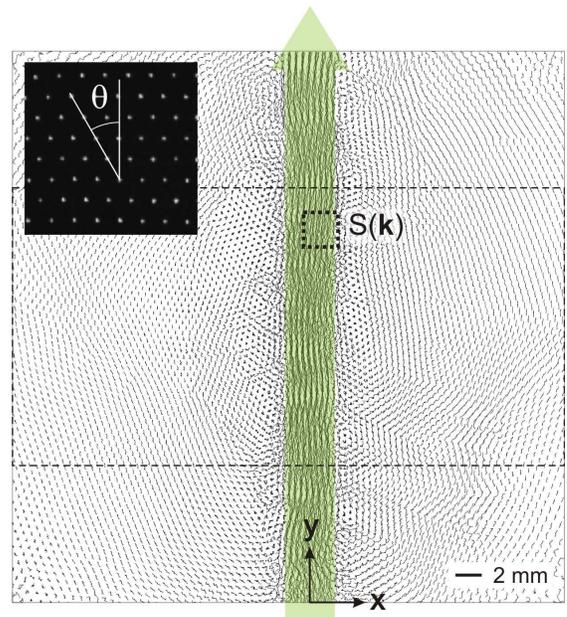}
\caption {\label {exp_layout} (color online) Shear stress is applied to a two-dimensional plasma crystal by the radiation pressure force of a rastered laser beam. The laser force is applied in the stripe $-2~{\rm mm} \leq x \leq 2~{\rm mm}$ in the positive $y$ direction. Particle trajectories in the camera field of view are shown during $1.67$~s of steady-state shear flow (corresponds to $P_{\rm laser}=2.4$~W in Figs.~\ref{vy_t},\ref{y_t}). Analysis was performed in the region indicated by the dashed-line rectangle, $S({\bf k})$ was calculated in the smaller area indicated by the dotted-line square. Before the laser was switched on, the plasma crystal was rotated so that the closely packed rows of particles formed an angle $\theta=30^{\circ}$ (see inset) or $0^{\circ}$ with the laser beam.
}
\end{figure}

Our experimental setup was a modified GEC (Gaseous Electronics Conference) rf reference cell \cite{Nosenko:11PRL}. Plasma was produced using a capacitively coupled rf discharge in argon at $0.66$~Pa. A single layer of dust particles was suspended in the plasma sheath of the lower rf electrode. The microspheres made of melamine formaldehyde had a diameter of $9.19\pm0.09$~$\mu$m, a mass $m=6.15 \times 10^{-13}$~kg, and acquired an electric charge of $Q=-16~000\pm1600e$ \cite{footnote1}. The suspension included around $9000$ particles and had a diameter of $\approx60$~mm. The mean interparticle distance in the center was $\Delta=0.51$~mm [measured from the first peak of the pair correlation function $g(r)$]. The Wigner-Seitz radius was $a\equiv(\pi n)^{-1/2}=0.27$~mm, where $n$ is the areal number density of particles. This corresponds to the screening parameter $\kappa \equiv a/\lambda_D=0.5$, where $\lambda_D$ is the screening length \cite{footnote1}. The neutral gas damping rate was $\nu=0.77~{\rm s}^{-1}$.

In our experimental conditions, the particle suspension self-organized in a highly ordered triangular lattice, as indicated by the high value of the (local) coupling parameter ${\it \Gamma} \equiv Q^2(4\pi\epsilon_0ak_BT)^{-1} \simeq 9000$, where $T$ is the particle kinetic temperature. When this plasma crystal settled, we rotated it to align one of the closely packed rows at $\theta=30^{\circ}$ or $0^{\circ}$ with respect to the laser beam, see inset in Fig.~\ref{exp_layout}. To rotate the crystal, we temporarily placed a small permanent magnet on the upper glass window \cite{Konopka:2000}. After the crystal was properly aligned, the magnet was removed.

Shear stress was applied to the plasma crystal by pushing a stripe of particles by the radiation pressure force of a rastered laser beam, as shown in Fig.~\ref{exp_layout}. A Millennia~PRO~15sJ solid-state $532$~nm laser with a maximum CW output power of $15$~W was used. The laser beam coming at a grazing angle of $\simeq 8^\circ$ was focused on the particle suspension to a spot smaller than the interparticle distance and then rapidly scanned to draw a rectangular stripe on the suspension. The scanning was performed by galvanometer-driven mirrors oscillating at a frequency of around $300$~Hz. The particles within the illuminated stripe reacted to the average radiation pressure force and were pushed in the direction of the laser beam. Shear stress was thus created in the particle suspension, its magnitude was controlled by varying the laser output power.

The lattice layer was illuminated by a $660$~nm, $20$~mW diode laser, with its output expanded into a horizontal sheet by a cylindrical lens. A video camera (Photron FASTCAM 1024 PCI) was mounted above the chamber, capturing a top view of the particle suspension with a size of $42.7\times42.7~{\rm mm}^2$ and resolution $1024\times1024~{\rm pixel}^2$. The camera lens was equipped with a narrow-band interference filter to admit only the illumination laser light scattered off the particles. The recording rate was set at $60$ frames per second.

\section {Results}

We applied increasing levels of shear stress to the crystal. At low stress, the crystal deformed elastically. At higher stress, the following stages were observed: defect generation while in a solid state, onset of plastic flow, and fully developed shear flow. In this paper, we will study in detail the last two stages.

\subsection {Development of shear-induced melting}

\begin{figure}
\centering
\includegraphics[width=0.9\linewidth]{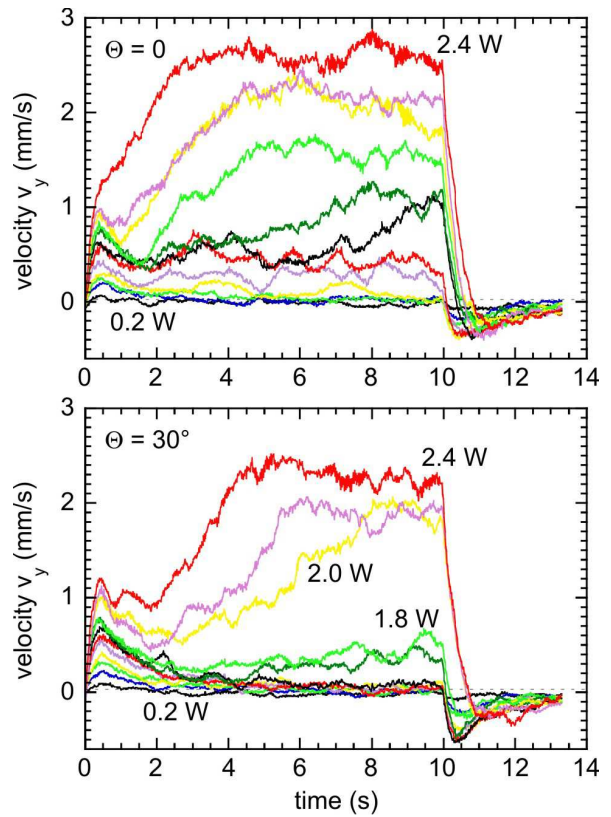}
\caption {\label {vy_t} Averaged particle velocity $V_y$ in the laser-illuminated stripe as a function of time, for two principal orientations of the plasma crystal. Different curves are for the manipulation laser power $P_{\rm laser}$ ranging from $0.2$~W to $2.2$~W, in increments of $0.2$~W. The laser was switched on at $t=0$ and off at $t=10$~s.}
\end{figure}

\begin{figure}
\centering
\includegraphics[width=0.9\linewidth]{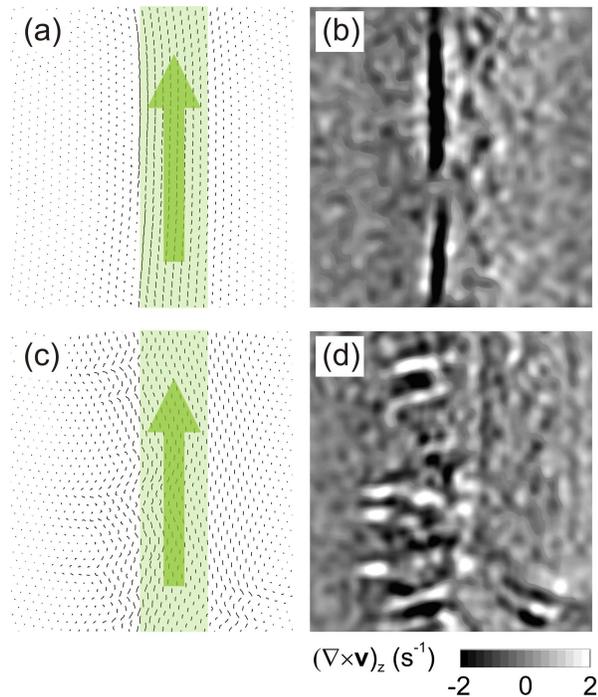}
\caption {\label {disl} (color online) (a),(c) Particle trajectories during $0.23$~s and (b),(d) the lattice vorticity $(\nabla \times \bf{v})_z$, where $\bf{v}$ is the velocity of individual particles, at the onset of plastic deformation. The particles in the shaded area were pushed by the laser ($P_{\rm laser}=2.4$~W), as indicated by the arrows. The lattice orientation before the laser was applied was (a),(b) $\theta=0^{\circ}$ and (c),(d) $\theta=30^{\circ}$. The lattice vorticity reveals shear motion and is used here to visualize moving dislocations \cite{Nosenko:07PRL}.}
\end{figure}

\begin{figure}
\centering
\includegraphics[width=0.9\linewidth]{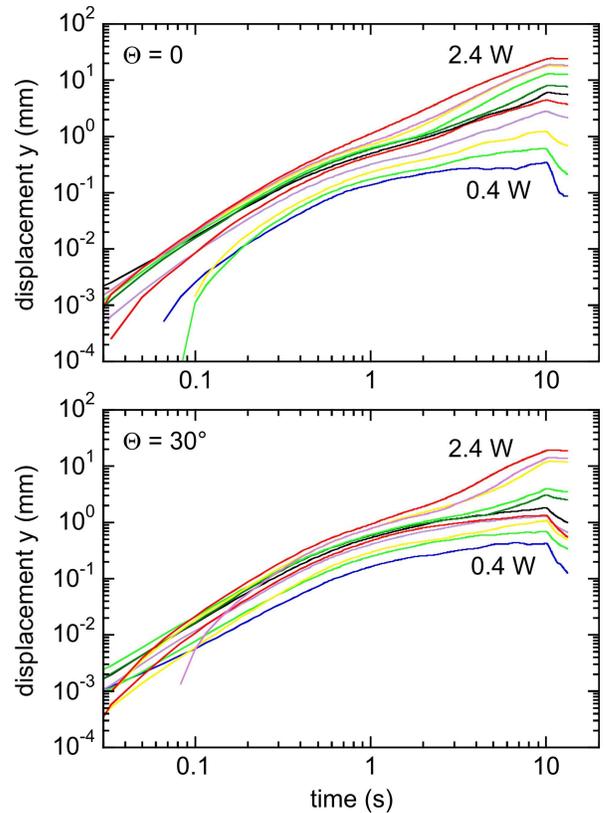}
\caption {\label {y_t} Displacement of the laser-illuminated particle stripe as a function of time, for two principal orientations of the triangular lattice. Different curves are for the manipulation laser power $P_{\rm laser}$ ranging from $0.4$~W to $2.2$~W, in increments of $0.2$~W. The laser was switched on at $t=0$ and off at $t=10$~s.}
\end{figure}

For an overview of the succession of stages leading to shear melting, we measured the average velocity $V_y$ \cite{footnote2} of all particles in the laser-illuminated stripe as a function of time, see Fig.~\ref {vy_t}. For $\theta=0^{\circ}$ (upper panel), there is a rather gradual change in the $V_y(t)$ curves for increasing laser power. The particle suspension becomes liquid and flowing at higher laser powers, but there is no evident threshold for shear melting.  In contrast, for $\theta=30^{\circ}$ (lower panel) the shear melting has a pronounced threshold character: By increasing the laser output power from $1.8$~W to $2.0$~W, the $V_y(t)$ curve changes dramatically. For laser powers $P_{\rm laser}\geq2.0$~W, the applied stress apparently exceeds the yield stress and the lattice eventually shear-melts. For $P_{\rm laser}=1.6-1.8$~W, the lattice defects (mostly dislocation pairs) are continuously generated which however do not lead to shear melting on the time scale of the present experiment. The finite values of $V_y$ in this case reveal a kind of a creep flow in the lattice \cite{Hartmann:Poster}.


Shear melting occurs more easily when closely packed rows of the crystal are aligned with the laser beam ($\theta=0^{\circ}$). This configuration favors dislocation nucleation and subsequent motion, since they glide in the direction of the laser beam \cite{Nosenko:11PRL}, see Figs.~\ref {disl}(a),(b). For $\theta=30^{\circ}$, the plastic deformation of the lattice occurs via dislocations gliding at a certain angle with respect to the laser beam, see Figs.~\ref{disl}(c),(d). Therefore, the melting happens at higher $P_{\rm laser}$ and has a more pronounced threshold character.


The first peak at $t\simeq0.5$~s in Fig.~\ref {vy_t} corresponds to in-cage oscillations of particles. Its presence is a manifestation of a gradual crossover from elastic to plastic deformation. It is naturally more pronounced for $\theta=30^{\circ}$.

We measured the displacement of the laser-illuminated particle stripe by integrating the average velocity $V_y$ \cite{footnote2} of all particles in the stripe, see Fig.~\ref {y_t}. After the laser is switched on at $t=0$ the displacement scales as $\propto t^2$, which indicates that the particles accelerate due to the action of the applied force. After $\simeq0.5$~s, the particle motion slows down. As long as $P_{\rm laser}$ is below a threshold the curves tend to saturate, whereas above the threshold the lattice shear-melts. Then $y\propto t^{\alpha}$, where $\alpha=0.9-1.7$, depending on the laser power. Notice the final push-back, even of a liquid complex plasma after the laser was switched off at $t=10$~s. This is caused by the elastic response of unmelted lattice and by the pressure gradient in the liquid state.

\subsection {Shear-induced reordering}

\begin{figure}
\centering
\includegraphics[width=0.85\linewidth]{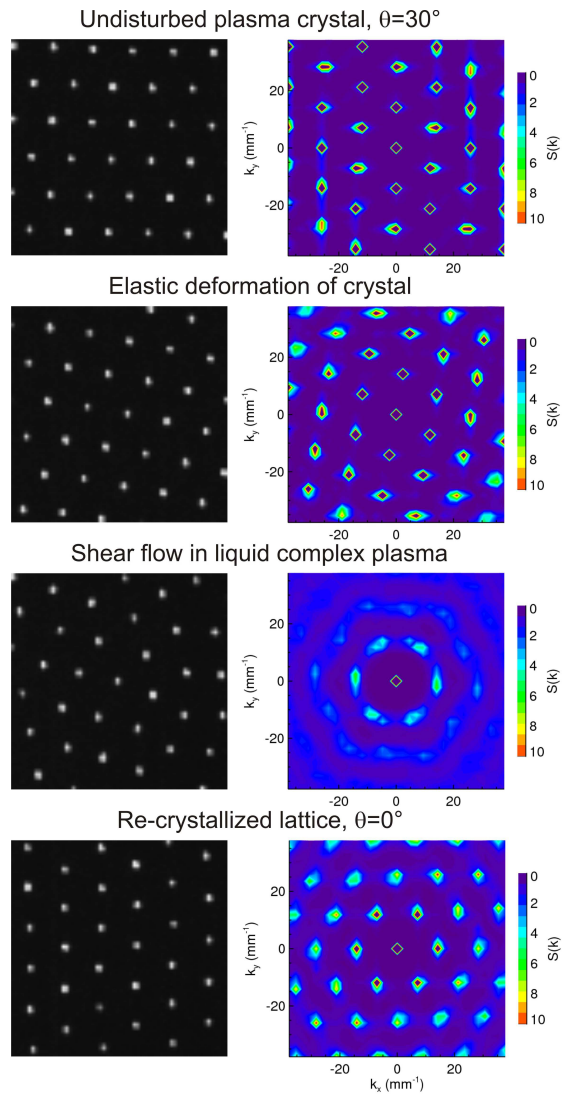}
\caption {\label {S_k} Shear-induced reordering in a 2D complex plasma. The left column shows snapshots of particle positions, the right column shows corresponding static structure factor $S({\bf k})$ (all from the area indicated by the dotted-line square in Fig.~\ref{exp_layout}). Undisturbed triangular lattice was initially oriented at $\theta=30^{\circ}$, first row. The lattice was distorted elastically $t=0.67$~s after the laser was applied ($P_{\rm laser}=2.4$~W), second row. During the subsequent steady-state shear flow, the complex plasma was in a liquid state which is, however, not completely isotropic and has a nascent triangular order with $\theta=0^{\circ}$, third row [$S({\bf k})$ averaged over $t=5-10$~s]. After the laser was switched off, the lattice re-crystallized with $\theta=0^{\circ}$, last row [$S({\bf k})$ averaged over $t=10-13.3$~s].
}
\end{figure}

We observed shear-induced reordering in a flowing liquid complex plasma. We made two related observations. First, a flowing liquid complex plasma is not completely isotropic. Rather, on top of isotropic particle distribution appears a weak triangular order with $\theta=0^{\circ}$ even in the cases when the original triangular lattice was oriented at $30^{\circ}$ (see Fig.~\ref {S_k}, third line). Similar anisotropy in a liquid complex plasma was observed in Ref.~\cite{Nosenko:12PRL}, where the initial triangular order with $\theta=0^{\circ}$ was not completely destroyed in a shear melted state. Second, after the laser was switched off, the liquid complex plasma re-crystallized in a triangular lattice with $\theta=0^{\circ}$ (see Fig.~\ref {S_k}, last line), even though the lattice was oriented at $\theta=30^{\circ}$ before melting.

A flowing liquid complex plasmas, therefore, favors particle ordering where ``closely packed lines'' are oriented along the flow. It was further observed that even outside of the shear flow the adjacent crystalline regions gradually change their orientation from $\theta=30^{\circ}$ to $0^{\circ}$ and stay in that orientation after cessation of the flow.

The observed particle ordering is similar to the formation of particle strings aligned in the flow direction in sheared 2D colloidal dispersions \cite{Stancik:2004} and simulated liquid of soft disks \cite{Butler:1996}. Here, we observed it for the first time in a sheared complex plasma. The underlying mechanism might be similar to the phenomenon of lane formation observed in binary systems where particles of one sort were driven through a cloud of particles of another sort, e.g. in colloidal dispersions \cite{New_book,Leunissen:05,Vissers:11} and complex plasmas \cite{New_book,Suetterlin:09,Du:12}.

\subsection {Rheology of liquid complex plasma}

\begin{figure}
\centering
\includegraphics[width=0.95\linewidth]{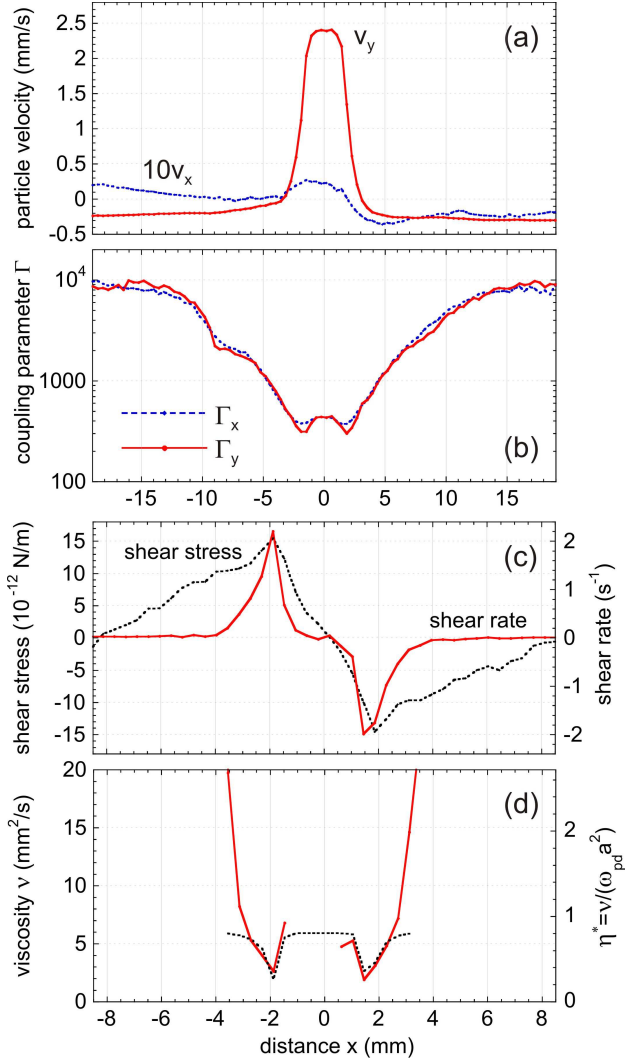}
\caption {\label {viscosity} (a) Averaged particle velocity $v_{x,y}$, (b) coupling parameter ${\it \Gamma}_{x,y}$, (c) shear stress $\sigma_{xy}$ and shear rate $\dot{\gamma}=\partial v_y/ \partial x$, and (d) kinematic viscosity $\nu=\rho^{-1}\sigma_{xy}/\dot{\gamma}$ (where $\rho=mn$ is the areal mass density) as functions of the transverse coordinate in the steady-state shear flow. The dimensionless viscosity $\eta^*=\nu/(\omega_{pd}\,a^2)$ is also shown, where $\omega_{pd}=[Q^2/2\pi\epsilon_0ma^3]^{1/2}$ is the relevant plasma frequency and $a=(\pi n)^{-1/2}$ is the Wigner-Seitz radius. The averaging was performed in narrow rectangular bins with the width of $\Delta x=0.42$~mm during $5.7$~s. The dotted line in (d) is a fit of experimental data with $\nu=\nu_0(1-\alpha\dot{\gamma}^2)$, where $\nu_0=5.92~{\rm mm}^2/{\rm s}$ and $\alpha=0.14~{\rm s}^2$. The lattice was oriented at $\theta=30^{\circ}$ before the laser was switched on, the laser power was $P_{\rm laser}=2.4$~W.
}
\end{figure}

In this section, we discuss the rheology of a steady-state shear flow in liquid complex plasma. The time $t_w$ for the flow to reach steady state varied depending on the initial crystal orientation and the laser power applied, see Fig.~\ref {vy_t}. For example, $t_w\simeq4$~s for $\theta=30^{\circ}$ and $P_{\rm laser}=2.4$~W. For this case, the particle trajectories in a steady-state shear flow are shown in Fig.~\ref{exp_layout}. The laser force is applied in the stripe $-2~{\rm mm} \leq x \leq 2~{\rm mm}$. Particles far from the flow are only slightly disturbed and the original crystalline structure here is intact. There is a boundary layer between the flow and unmelted crystal. Profiles of time-averaged components of the particle velocity $v_{x,y}(x)$ are shown in Fig.~\ref{viscosity}(a), and the longitudinal shear strain rate $\dot{\gamma}=\partial v_y/ \partial x$ is plotted in Fig.~\ref{viscosity}(c) (solid curve). The profiles of the coupling parameter are shown in Fig.~\ref{viscosity}(b), where ${\it \Gamma}_{x,y}$ corresponds to the particle kinetic temperature $T_{x,y}$ \cite{footnote3}. The values of ${\it \Gamma}_x \simeq {\it \Gamma}_y \approx 9000$ far from the flow indicate a highly ordered crystal. There is a small anisotropy in the coupling parameter, in particular at the edges of the flow, where ${\it \Gamma}_y \lesssim {\it \Gamma}_x$, indicating noticeable deviation of the system (driven in the $y$ direction) from equilibrium.

We note that the values of coupling parameter ${\it \Gamma}_y \simeq {\it \Gamma}_x \approx 430$ in the middle of the flow and ${\it \Gamma}_y \approx 300$ at its edges are still above the {\it equilibrium melting line} in the $(\kappa, {\it \Gamma})$ coordinates as reported in simulations \cite {Hartmann:2005,Vaulina:2002}.
This is because melting in our experiment is essentially a non-equilibrium phenomenon driven by external forces. Indeed, by comparing the relative flow velocity $v_{\rm flow} \simeq \dot{\gamma}\Delta$ with the thermal velocity at the flow edges, $v_{T_y}=\sqrt{k_BT_y/m}$, we conclude that these are about the same. Therefore, the effective ``non-equilibrium'' coupling parameter (in the $y$ direction), which is inversely proportional to $v_{\rm flow}^2+v_{T_y}^2$, is substantially reduced with respect to the ``equilibrium'' value of ${\it \Gamma}_y$.

To characterize the flow, we calculate the shear stress $\sigma_{xy}$ in the particle suspension using the positions of individual particles, as explained below. We used a method where the stress at the position of a particle is built up from individual contributions of its neighbors. In this approach, the stress tensor is given by \cite{New_book}:
\begin{equation}
\label{stress1}
{\mathbf \Sigma}=-\frac12 n^2 \int d{\mathbf r} \frac{{\mathbf r} \otimes {\mathbf r}}{r} \frac {dV}{dr} g({\mathbf r}),
\end{equation}
where $g({\mathbf r})$ is pair correlation function and $V(r)$ is the pair interaction potential. Then for our 2D system we obtain in polar coordinates:
\begin{equation}
\label{stress2}
\sigma_{xy}=-\frac12 n^2 \int_0^{r_{\rm max}} dr \, r^2 \frac {dV}{dr} \int_0^{2\pi} d\phi \cos \phi  \sin \phi \, g(r,\phi).
\end{equation}
To calculate $\sigma_{xy}$ from Eq.~(\ref{stress2}), $V(r)$ was approximated by the Yukawa potential and $16$ shells of neighbors were included in $g(r,\phi)$. The result is shown in Fig.~\ref{viscosity}(c) (dotted curve). Note that we calculate only the potential part of shear stress, since the remaining kinetic part is negligible in our experimental conditions (${\it \Gamma} > 300$) \cite{Liu:2005}.

\begin{figure}
\centering
\includegraphics[width=0.9\linewidth]{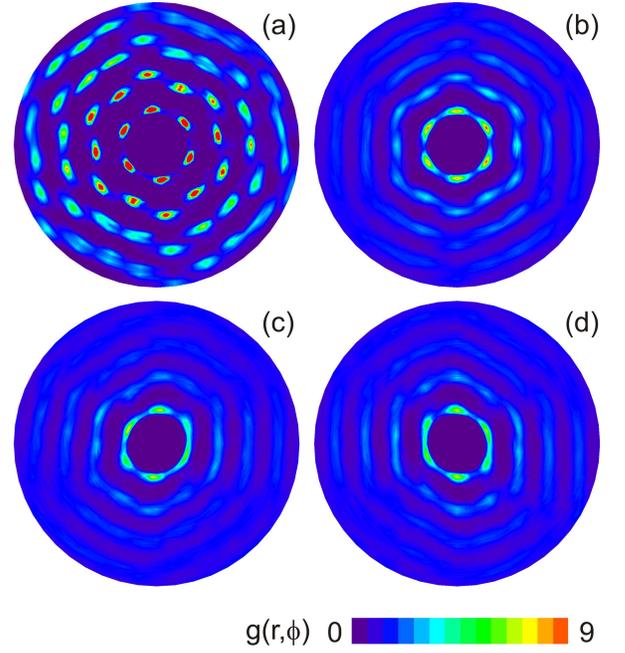}
\caption {\label {g_r_phi} Pair correlation function $g(r,\phi)$ of complex plasma (a) in a crystalline area far from the shear flow, $x=8.1$~mm, (b) in the middle of the shear flow, (c),(d) at its edges, $x=-1.9$~mm and $1.9$~mm, respectively (see Fig.~\ref{viscosity}). The peak amplitudes are up to $41$ in (a) and therefore appear saturated in this color coding. Elliptically distorted $g(r,\phi)$ in (c),(d) indicates strong shear stress. Only $4$ shells of neighbors are shown here, shear stress was calculated using $16$ shells.}
\end{figure}

It is clear from Eq.~(\ref{stress2}) that the shear stress is determined by the $\sin 2\phi$ component of $g(r,\phi)$. As it was shown in Ref.~\cite{Nosenko:12PRL}, the interparticle spacing $\Delta(\phi)$ [representative of the first peak of $g(r,\phi)$] in a sheared liquid complex plasma indeed has a strong $\sin 2\phi$ component. In Fig.~\ref{g_r_phi}, we present the complete $g(r,\phi)$, where the $\sin 2\phi$ component shows up as an elliptical distortion particularly clearly seen in the first peak, see panels (c) and (d).

Given the shear stress and strain rate, the kinematic viscosity of liquid complex plasma was calculated as $\nu=\rho^{-1}\sigma_{xy}/\dot{\gamma}$, where $\rho=mn$ is the areal mass density. The result for $\nu$ is shown in Fig.~\ref{viscosity}(d) (solid curve). Shear viscosity has two local minima at the edges of the flow, $x=-1.9$~mm and $1.5$~mm, where the shear rate is maximal. This indicates shear thinning, i.e., a kind of non-Newtonian behavior of a liquid whose viscosity diminishes for higher strain rates. Shear thinning in complex plasmas was demonstrated in experiments \cite{Ivlev:07} and simulations \cite{Donko:2006}. In the region $-1.4~{\rm mm}<x<0.6~{\rm mm}$, where $\dot{\gamma} \simeq 0$ the shear viscosity cannot be calculated as $\sigma_{xy}/\dot{\gamma}$ and therefore is not shown in Fig.~\ref{viscosity}(d).

The found values of kinematic viscosity compare well with the results of Ref.~\cite{Nosenko:04PRL_visc}. In the present work, $\nu \simeq 2.5~{\rm mm}^2/{\rm s}$ at the edges of shear flow where ${\it \Gamma}=300-370$. In Ref.~\cite{Nosenko:04PRL_visc}, $\nu \simeq 2~{\rm mm}^2/{\rm s}$, for similar values of ${\it \Gamma}$ and $\kappa$. In both experiments, shear thinning was apparently well pronounced. In Ref.~\cite{Nosenko:04PRL_visc}, viscosity was calculated from the fits of experimental particle velocity profiles to the solution of a modified Navier-Stokes equation taking neutral gas drag into account. It is remarkable that the two very different methods of calculating shear viscosity give similar results.

\begin{figure}
\centering
\includegraphics[width=0.95\linewidth]{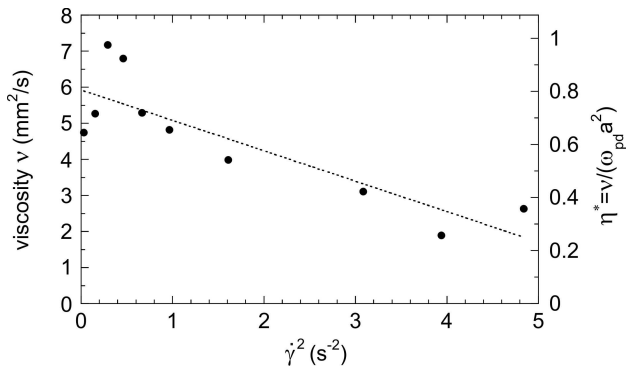}
\caption {\label {nu_gammagot2} Kinematic viscosity $\nu$ is shown as a function of $\dot{\gamma}^2$, revealing shear thinning. The line is a fit of experimental data with $\nu=\nu_0(1-\alpha\dot{\gamma}^2)$, where $\nu_0=5.92~{\rm mm}^2/{\rm s}$ and $\alpha=0.14~{\rm s}^2$. The dimensionless viscosity $\eta^*=\nu/(\omega_{pd}\,a^2)$ is also shown.}
\end{figure}

To characterize shear thinning, we plot kinematic viscosity $\nu$ as a function of $\dot{\gamma}^2$, see Fig.~\ref{nu_gammagot2}. Viscosity declines roughly three times when $\dot{\gamma}^2$ increases from $0$ to $5~{\rm s}^{-2}$. Note that simultaneously the coupling parameter ${\it \Gamma}$ varies by $14-42\%$, see Fig.~\ref{viscosity}(b), due to the viscous heat release \cite{Nosenko:08PRL} and higher temperature at the edges of the flow, where the shear rate is higher. The variations in $\dot{\gamma}$, ${\it \Gamma}$, and $\nu$ are self-consistent due to the dependence of shear viscosity on temperature \cite{Ivlev:07}. The value of viscosity in the limit of low shear, $\nu_0=5.92~{\rm mm}^2/{\rm s}$ can be compared to the results of equilibrium simulations \cite{Liu:2005,Budea:2012}. It corresponds to dimensionless viscosity $\eta^*\equiv\nu/(\omega_{pd}\,a^2)=0.8$, where $\omega_{pd}=[Q^2/2\pi\epsilon_0ma^3]^{1/2}$ and $a=(\pi n)^{-1/2}$. In the equilibrium 2D MD simulation of Ref.~\cite{Liu:2005}, where viscosity was calculated using the Green-Kubo formula, $\eta^*=0.5-1.6$ for the highest reported values of ${\it \Gamma}=125-130$, for a comparable value of $\kappa$.

\section {Summary}

Shear induced melting of a 2D plasma crystal was studied experimentally. Shear stress was applied by a focused laser beam rastered over a rectangular area in the crystal. For the first time, the study was angle-resolved, i.e., the closely packed rows of the crystal were oriented at $\theta=0^{\circ}$ or $30^{\circ}$ (two principal orientations) with respect to the shearing force. Shear melting was found to be strongly anisotropic. The transition of the crystal to flowing liquid with increasing shear stress was rather gradual for $\theta=0^{\circ}$ and had a distinctly threshold character for $30^{\circ}$. This is explained by the role of dislocations in the crystal plastic deformation. Their nucleation and motion was facilitated for $0^{\circ}$ and retarded for $30^{\circ}$.

The rheology of the steady-state shear flow in liquid complex plasma was studied. Shear stress and strain rate were measured with spatial resolution and from these, the shear viscosity was calculated. Shear viscosity had minima in the regions with high velocity shear, demonstrating shear thinning.

Shear-induced reordering was observed in flowing liquid complex plasma, where particles tended to form strings aligned in the direction of the flow. Different from laning in driven binary complex plasmas, formation of strings in sheared complex plasma was observed for the first time.

\acknowledgments

We thank J\"{u}rgen Horbach for valuable discussions.


\begin{thebibliography}{}
\bibitem {footnote2} To take into account the return flow of particles at the suspesion perifery, $V_y$ was calculated in every frame as $V_y=\langle v_y\rangle _{_{\rm stripe}}-\langle v_y\rangle _{_{\rm fov}}$, where the averaging was performed respectively for all particles in the laser-illuminated stripe and in the complete field of view. The average $V_x$ component was on the scale of $1$\% of $V_y$.
\bibitem {footnote3} Particle kinetic temperature was calculated as $T_{x,y}=m(\langle v_{x,y}^2\rangle _{_t}-\langle v_{x,y}\rangle ^2 _{_t})/k_B$, where the averaging was performed over time.
\end{thebibliography}
\end{document}